\documentclass[notitlepage,aps,epsfig,showpacs,floats,twocolumn,amssymb,amsmath,floatfix,
groupedaddress,superscriptaddress]{revtex4-1}
\usepackage{graphicx} 
\usepackage{subfigure}
\usepackage{dcolumn} 
\usepackage{bm} 
\usepackage{float} 
\usepackage{bbm} 
\usepackage{enumitem,xcolor}
\usepackage{color}
\usepackage{amsfonts}
\usepackage{xcolor}
\usepackage{lmodern} 
\usepackage{natbib}
\usepackage{comment}
\usepackage{amsmath} 
\usepackage{amsmath,amssymb}
\usepackage{natbib}
\usepackage[margin=2cm]{geometry}
\usepackage{tikz}
\usepackage{ulem}
\usetikzlibrary{matrix}

\newcommand\bi{\begin{itemize}}
\newcommand\ei{\end{itemize}}

\usepackage[colorlinks=true,citecolor=blue]{hyperref}
\hypersetup{colorlinks=true,citecolor=blue,linkcolor=blue,urlcolor=black}

    \usepackage{geometry}

    \usepackage{tikz} 
    \usetikzlibrary{shapes,arrows,calc}
    \usetikzlibrary{datavisualization}
    \usetikzlibrary{datavisualization.formats.functions}
    \usepackage{pgfplots} 
    \usepackage{tikz}
\usetikzlibrary{matrix}  

\usepackage[colorlinks=true,citecolor=blue]{hyperref}

\begin{document}
\title{Half-integer and full-integer topological defects in polar active matter:\\ Emergence, crossover, and coexistence}
\author{Aboutaleb Amiri}

	\affiliation{Max Planck Institute for the Physics of complex systems, Dresden, Germany}
	
		\author{Romain Mueller}
	\affiliation{Rudolf Peierls Centre for Theoretical Physics, University of Oxford, UK}
	
	\author{Amin Doostmohammadi}
	\email{doostmohammadi@nbi.ku.dk}
	\affiliation{The Niels Bohr Institute, University of Copenhagen, Copenhagen, Denmark}

\begin{abstract}
The presence and significance of active topological defects is increasingly realised in diverse biological and biomimetic systems. We introduce a continuum model of polar active matter, based on conservation laws and symmetry arguments, that recapitulates both polar and apolar (nematic) features of topological defects in active turbulence. Using numerical simulations of the continuum model, we demonstrate the emergence of both half- and full-integer topological defects in polar active matter. Interestingly, we find that crossover from active turbulence with half- to full-integer defects can emerge with the coexistence region characterized by both defect types. These results put forward a minimal, generic framework for studying topological defect patterns in active matter which is capable of explaining the emergence of half-integer defects in polar systems such as bacteria and cell monolayers, as well as predicting the emergence of coexisting defect states in active matter. 
\end{abstract}
\maketitle

Topological defects denote singularities in the order parameter field, marking the regions where the order breaks down~\cite{kosterlitz2016kosterlitz,michel1980symmetry,bowick2009two}. They are topological in the sense that no smooth local variation in the order parameter space can remove them~\cite{de1993physics,chaikin_lubensky_1995} and they are prevalent in various physical systems ranging from cosmic strings in particle physics model of early universe~\cite{durrer2002cosmic} to vortices in superfluid helium films~\cite{kosterlitz2016kosterlitz} and flux tubes in superconductors~\cite{tinkham2004introduction}, to disclination lines in liquid crystals~\cite{poulin1997novel}. More recently, topological defects are being increasingly identified in biological systems, where, in analogy to liquid crystals, they mark singularities in the orientation field associated with the alignment of the constituents of biological systems. These range from topological defects in subcellular filaments such as actin~\cite{schaller2013topological,kumar2018tunable} or microtubules~\cite{nedelec1997self,roostalu2018determinants,duclos2020topological}, to defects in bacterial alignment~\cite{dell2018growing,meacock2021bacteria}, and topological defects in the orientation field associated with the elongation of fibroblasts~\cite{duclos2017topological}, epithelial~\cite{saw2017topological}, and stem cells~\cite{kawaguchi2017topological}. Remarkably, not only these topological defects are found within numerous biological systems and across subcellular to multicellular scales, they appear to play an important role in various biological processes such as cell death and extrusion in epithelia~\cite{saw2017topological}, accumulation sites for bacteria and stem cells~\cite{meacock2021bacteria,copenhagen2020topological,kawaguchi2017topological}, and determinants of morphological features in bacterial colonies~\cite{doostmohammadi2016defect} and developing animal {\it Hydra}~\cite{maroudas2020topological}. The distinguishing feature of these realizations of topological defects, compared to their counterparts in non-living systems, is that they are characterized by active flows continuously being generated due to the activity of the constituents of the living material (see~\cite{doostmohammadi2018active,shankar2020topological} for recent reviews of active topological defects).

Despite abundant and growing identifications of topological defects in various living biological systems, fundamental questions regarding their nature remain unanswered. One particularly puzzling observation is the abundance of half-integer defects (defects with nematic, head-tail, symmetry) in systems with a clear polar symmetry such as motile bacteria~\cite{copenhagen2020topological,meacock2021bacteria} and eukaryotic cells~\cite{saw2017topological,kawaguchi2017topological,blanch2018turbulent}. Polarity in this context determines the direction of motion of the self-propelled cell and yet when multicellular collections of these polar entities are formed, such as in biofilm layers or epithelial monolayers, the emerging topological defects show half-integer charges - in systems with polar symmetry full-integer defects are expected~\cite{meyer1973existence,kruse2004asters}. 

Indeed, continuum theories of polar active matter predict topological defects in the form of asters, vortices and spirals that have a full-integer charge~\cite{kruse2004asters,elgeti2011defect} as has been reported for microtubule-motor protein mixtures~\cite{nedelec1997self,surrey2001physical}. On the other hand, based on the emergence of half-integer defects in several cellular systems, most of existing continuum models treat these as active nematics, describing the coarse-grained alignment of constituent particles through nematic tensor, modeling activity through a stress term proportional to this nematic tensor, and neglecting polarity altogether. It is not clear how these distinct, yet relevant, symmetries compete in a real biological system, nor why half-integer defects emerge in active polar systems. While recent experiments and agent-based models have reported coexisting polar waves and nematic bands in actomyosin motility assay~\cite{huber2018emergence}, the study of topological defects in systems with mixed polar-nematic symmetry is non-existent and a generic theoretical framework for describing defect dynamics in such systems is lacking.

Here we address this problem by introducing a continuum formulation for polar active entities that shows the emergence of both half-integer and full-integer defects within one framework. Moreover, compared to the current models of active nematics that are based on tensorial equations for the nematic orientation, we introduce a vector-based model that holds all the essential features and allows for recovering both polar and nematic topological defects.\\
\begin{figure*}[th]
\centering
  \includegraphics[width=1.0\linewidth]{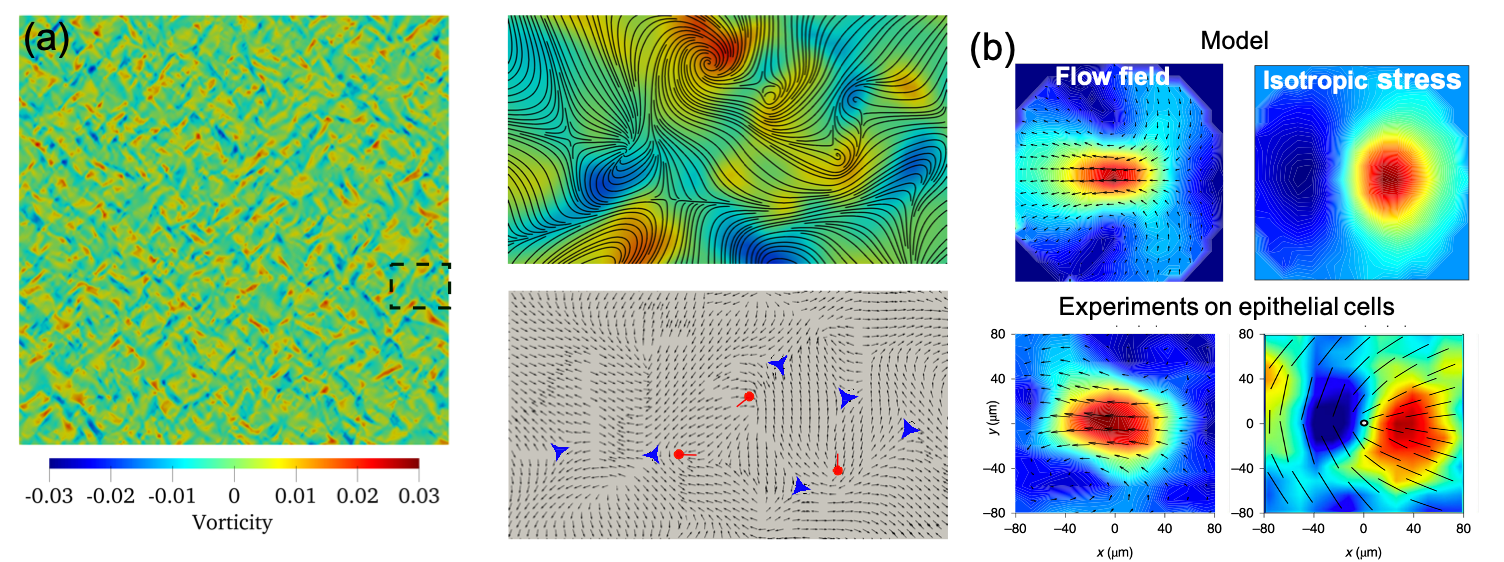}
  \caption{(a) Snapshots of active nematic turbulence state from continuum equations of polar active matter with the elasticity ratio $K_p/K=0.0$. Shown here are the vorticity of the full computational domain ({\it left})  and a selected zoomed-in region marked by dashed lines ({\it right}). In the zoomed-in regions black solid lines ({\it top}) and black arrows ({\it bottom})  represent streamlines and the  polar director field, respectively. In the bottom zoomed-in panel the half-integer $+1/2$, $-1/2$ topological defects are marked by red comets and blue triangles, respectively. (b) Corresponding averaged flow field and isotropic stress around $+1/2$ topological defects from continuum equations of polar active matter with the elasticity ratio $K_p/K=0.0$ ({\it top}) and are compared to the experimental measurements for epithelial monolayers of MDCK cells ({\it bottom}). In the flow fields colormaps indicate velocity magnitude and black arrows indicate velocity vectors, while in the stress maps, the colormaps indicate isotropic stresses. All colormaps are normalized by maximum values such that the velocity magnitudes range between $0$ ({\it blue}) to $+1$ ({\it red}) and isotropic stress maps range between $-1$ ({\it blue}) to $+1$ ({\it red}). Experimental measurements courtesy of Lakshmi Balasubramaniam from Benoit Ladoux's lab.}
  \label{fig:nem}
\end{figure*}

We begin by describing a force balance equation governing both self-propulsion and active stress generation in the same active system. 
We describe the direction of self-propulsion by a local polarity vector $\vec{p}$ such that each active particle generates a polar force $\vec{f}^{\text{pol.}}=\alpha\vec{p}$, where $\alpha$ controls the force strength. In addition to this polar force, each particle generates a dipolar contribution to the active stress that can be described at leading order by the stresslet ${\boldsymbol \sigma}^{\text{active}}=-\zeta \left(\vec{p} \vec{p}^{\text{T}} - p^2{\bf I}/2 \right)$~\cite{prost2015active,julicher2018hydrodynamic}, where ${\bf I}$ is the identity tensor, and $\zeta$ is the activity coefficient. The active contributions, from the polar force and active stress, are balanced by viscous and elastic passive stresses and any existing friction with the underlying substrate, through the momentum equation
\begin{equation}
    \rho \left(\partial_t \vec{u} + \vec{u}\cdot\vec{\nabla}\vec{u} \right) = \vec{f}^{\text{pol.}} + \vec{\nabla}\cdot {\boldsymbol \sigma}^{\text{active}} + \vec{\nabla}\cdot {\boldsymbol \sigma}^{\text{passive}} - \xi \vec{u},\label{eq:u}
\end{equation}
where $\rho$ is the fluid density, $\vec{u}$ the velocity field that satisfies incompressibiliy condition ($\vec{\nabla}\cdot\vec{u}=0$), and $\xi$ the friction coefficient. The passive stress ${\boldsymbol \sigma}^{\text{passive}}$ comprises pressure, elastic stress ${\boldsymbol \sigma}^{\text{elastic}}$, and viscous stress ${\boldsymbol \sigma}^{\text{viscous}}=2\eta {\bf E}=2\eta(\vec{\nabla}\vec{u})^{\text{S}}$, where $\eta$ is the shear viscosity and ${\bf E}$ is the rate of strain tensor characterizing the symmetric part of the velocity gradient. The elastic stress ${\boldsymbol \sigma}^{\text{elastic}}=\frac{\lambda+1}{2}\vec{p}\vec{h} + \frac{\lambda-1}{2}\vec{h}\vec{p} - \frac{\lambda}{2}\vec{p}\cdot\vec{h}$, is described in terms of the polarity field and its conjugate field $\vec{h}$ known as the molecular field (defined below), with $\lambda$ denoting the flow alignment parameter~\cite{de1993physics}.
For completeness we have retained all inertial terms and passive stresses in Eq.~\eqref{eq:u}. In many biological realizations of active matter, including microscopic active particles, the inertia is negligible compared to strong viscous dissipation and as such the lhs of \eqref{eq:u} drops out. Furthermore, elastic stresses are normally dominated by active contributions and are commonly neglected in studies of active systems~\cite{giomi2014defect,blanch2017hydrodynamic}. Under these conditions the momentum equation simplifies to a force balance
\begin{equation}
    \xi \vec{u} - \alpha\vec{p} =  \eta\nabla^2\vec{u} - \zeta \vec{\nabla}\cdot \left(\vec{p} \vec{p}^{\text{T}} - p^2\frac{{\bf I}}{2} \right).\label{eq:usimple}
\end{equation}
The above force balance equation is only represented for simplicity. The overall framework and all the conclusions presented in this paper hold in the presence of inertial terms and elastic stresses as well.

The velocity field is coupled to the spatio-temporal evolution of the polarity vector through:
\begin{equation}
    \partial_t \vec{p} + \vec{u}\cdot\vec{\nabla}\vec{p} + \lambda {\bf E}\cdot \vec{p} + {\boldsymbol \Omega}\cdot \vec{p}  = \frac{1}{\gamma} \vec{h},\label{eq:p}
\end{equation}
where the flow alignment parameter $\lambda$ characterizes the response of the polar alignment to the symmetric and anti-symmetric parts of the velocity gradient tensor (denoted by the rate of strain ${\bf E}$ and vorticity ${\boldsymbol \Omega}$ tensors, respectively), and $\gamma$ is the rotational viscosity that controls the relaxation of the polar alignment to the minimum of the free energy $\mathcal{F}$ through the molecular field $\vec{h}=-\delta\mathcal{F}/\delta \vec{p}$. The free energy is described as
\begin{eqnarray}
     \mathcal{F}=\int d\vec{x} &\Bigg\{A\left(-\frac{p^2}{2}+\frac{p^4}{4}\right)~~~~~~~~~~~~~~~~~~~~~~~~~~~~\\ \nonumber
     &+\frac{K_p}{2}(\vec{\nabla}\vec{p})^2 + \frac{K}{2}\left(\vec{\nabla}(\vec{p}\vec{p}^{\text{T}}-p^2\frac{{\bf I}}{2})\right)^2\Bigg\},\label{eq:F}
\end{eqnarray}
where the first term under the integral controls the isotropic-polar transition favoring the emergence of finite polarity at $|\vec{p}|=1$. The second term penalizes gradients in polarity and is controlled by the elasticity coefficient $K_p$. However, by virtue of the symmetry additional term is allowed that penalizes gradients in nematic (apolar) alignment of the particles with its strength controlled by the coefficient $K$. 
Various additional contributions to the free energy have been considered in coarse-grained models of self-propelled particles introducing {\it separate} equations for polarity vector and nematic tensor fields~\cite{baskaran2008hydrodynamics,baskaran2010nonequilibrium,baskaran2012self,chate2019dry,vafa2020defect}. 
As we show here, the competition between the additional nematic elasticity, controlled by $K$, and the regular polar elasticity, controlled by $K_p$ allows for continuous transition between polar and nematic topological defects. As such the model presented here provides, generic, minimal continuum formulation of active self-propelled particles, with emergent polar and nematic properties.

We simulate equations~\eqref{eq:u},\eqref{eq:p} using a hybrid lattice-Boltzmann method, combining finite-difference method for the evolution of polarity vector Eq.~\eqref{eq:p}, and the lattice-Boltzmann method for solving the Navier-Stokes equation Eq.~\eqref{eq:u} with $\rho=40$ and $\eta=2/3$ in lattice Boltzmann units, ensuring that the Reynolds number in the simulations is negligible $Re \ll 1$~\cite{thampi2014instabilities,doostmohammadi2017onset}. Simulations are initialized by setting quiescent velocity field and random polar alignments and periodic boundary conditions are used on the domain of the size $L_x \times L_y = 512 \times 512$. The dimensionless ratio of the viscosities is fixed $\eta/\gamma=2/3$, the alignment parameter is $\lambda=0.1$, and the dimensionless activity $\bar{\zeta}=\zeta/A$ and elasticities $K_p/K$ are varied throughout this study.

We begin by investigating the case of pure apolar elasticity $K_p/K=0$, setting the polar elasticity to zero ($K_p=0$). At first, we also set the polar and friction to zero $\alpha=0$, $\xi=0$, which reduces the number of parameters while still allowing to retain the minimal physics that show topological defect formation and the decisive role of the new apolar elastic term. As shown in Fig.~\ref{fig:nem}a, at statistical steady-state, the system evolves into active turbulence characterized by chaotic patterns of polar ordering and flow vortices~\cite{Wensink2012,alert2020universal}. Remarkably, the orientation field associated with the polarity vectors demonstrates the emergence of half-integer topological defects, in the form of comet-shaped $+1/2$ and trefoil shaped $-1/2$ defects, interleaving the vorticity patterns in the velocity field. This is striking, since the emergence of such half-integer defects has so far been only associated to continuum active nematics where a nematic tensor is described to account for coarse-grained orientation of active particles~\cite{simha2002hydrodynamic,ramaswamy2010mechanics,Marchetti13,doostmohammadi2018active}. More importantly, measuring the average velocity field around the emergent half-integer defects confirms their self-propulsive feature and produces flow patterns in agreement with experimentally measured flow fields for twitching bacteria~\cite{meacock2021bacteria} and epithelial and progenitor stem cell layers~\cite{Saw2017,kawaguchi2017topological,balasubramaniam2020nature} (see Fig.~\ref{fig:nem}b). Additionally, the isotropic stress patterns characterizing the compressive and tensile stresses around the self-propelled defect are in agreement with the experimentally measured stresses for epithelial monolayers of Madine Darby Canine Kidney (MDCK) cells (Fig.~\ref{fig:nem}b). Together, these results show that, at the limit of zero polar elasticity $K_p=0$, our proposed minimal model reproduces the emergence and active dynamics of half-integer topological defects, from spatio-temporal evolution of the polarity vectors and without revoking any dynamic evolution for an additional nematic tensor.
\begin{figure}[t]
\centering
   {\includegraphics[width=1.0\linewidth]{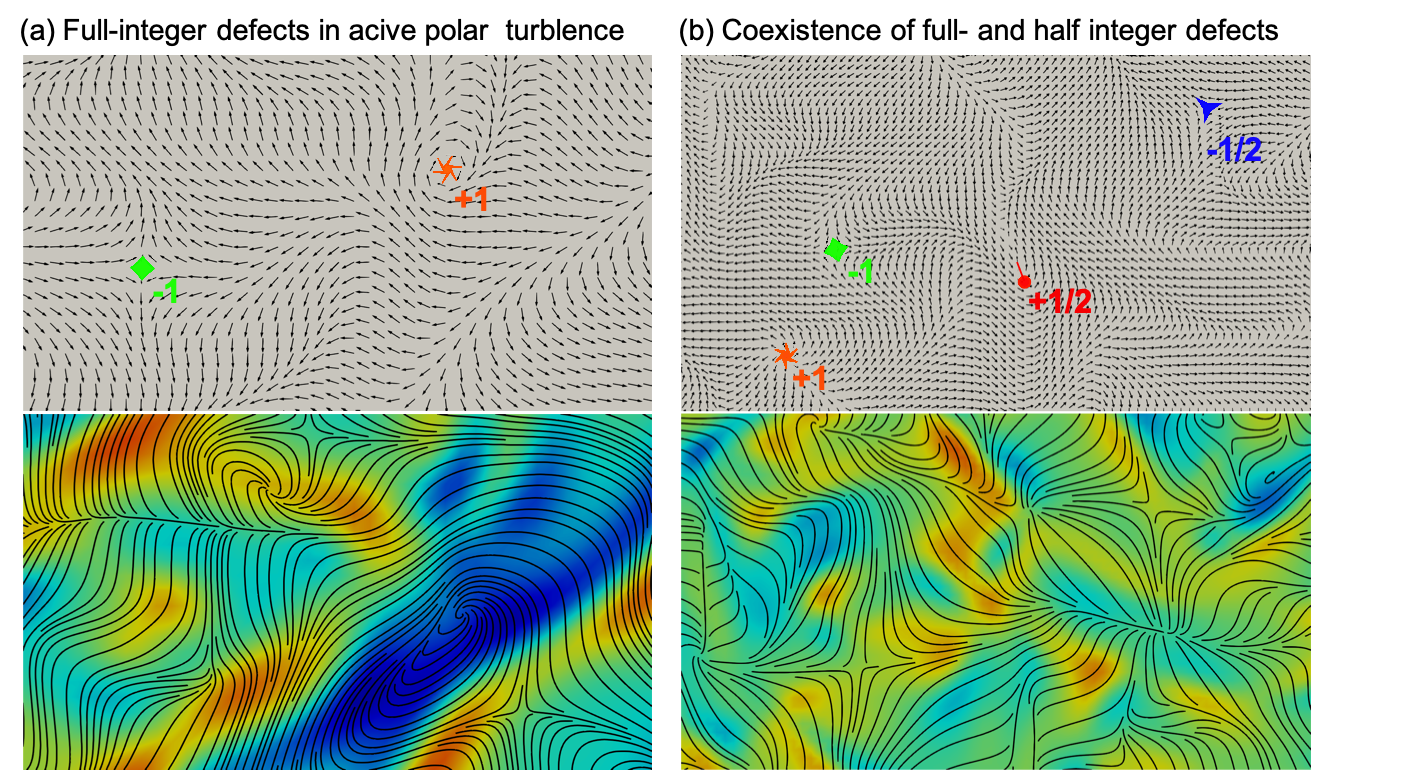}}
  \caption{{\bf Emergence of active polar turbulence and active turbulence state with mixed symmetry.} The colormaps are the same as in Fig.~1 and only a zoomed-in section of the entire simulation domain is shown. The full-integer $+1$, $-1$ topological defects are marked by orange asters and green squares, while the half-integer $+1/2$, $-1/2$ topological defects are marked by red comets and blue triangles, respectively.}
  \label{fig:pol}
\end{figure}

We find that at the other limit of zero apolar elasticity $K=0$ and finite polar elasticity $K_p/K=\infty$, the minimal model recovers the emergence of active turbulence interleaved by full-integer topological defects in the form of spiral vortex $+1$ and anti-vortex $-1$ defects (Fig.~\ref{fig:pol}a). The existence of such full-integer defects in active polar systems has been predicted theoretically~\cite{kruse2004asters} and shown experimentally in motility assays of actin or microtubule filaments where the motion is generated by motor proteins~\cite{surrey2001physical,schaller2013topological,ross2019controlling}. Numerical studies have explored the hydrodynamics of interactions between pairs of full-integer defects in active polar systems~\cite{elgeti2011defect}, however - surprisingly - the emergence and dynamics of active turbulence interleaved with full-integer defects is not well-understood. In particular, earlier studies of the active polar systems did not find any full-integer defects in the active turbulence phase~\cite{giomi2012polar} and only recently it has been shown numerically that such defects can mark the transition from active polar turbulence to phase turbulence upon increasing polarity~\cite{chatterjee2019fluid}.
\begin{figure}[t]
\centering
  \includegraphics[width=0.95\linewidth]{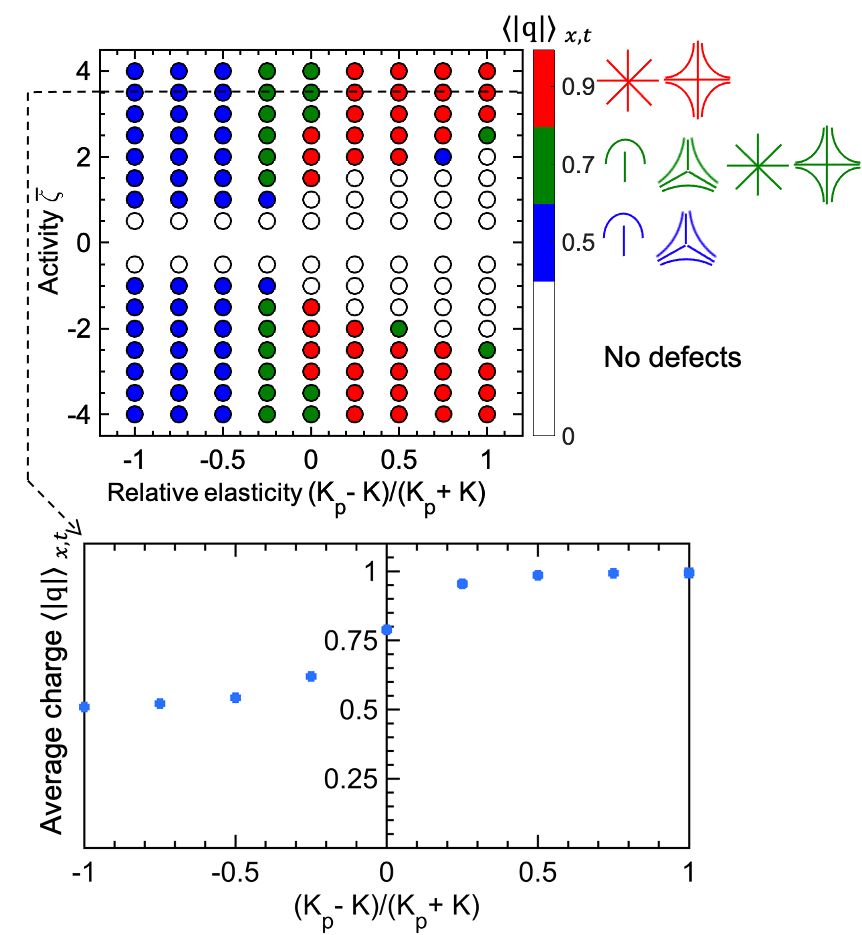}
  \caption{{\bf Stability-diagram in the activity-relative elasticity phase space.} Colormap indicates the average charge of the system with $0$ corresponding to no topological defect, $0.5$ to half-integer, and $1.0$ to full-integer topological defects, with values between $0.5-1.0$ marking the co-existence region. ({\it lower panel}) A cut in the phase space marked by dashed black line corresponding to activity value $\bar{\zeta}=3.5$.}
  \label{fig:charge}
\end{figure}

Importantly, when both polar and apolar elasticities are finite and non-zero, the minimal model predicts the emergence of an active topological state with a mixed symmetry. Here, the active turbulence, characterized by the emergence of vortices and jets in the flow profile, is accompanied by the emergence of both half-integer and full-integer topological defects in the orientation of polar director field. Therefore, varying the ratio of polar and apolar elasticities $K_p/K$ from a purely apolar elasticity ($K_p=0$) to a purely polar elasticity ($K=0$) results in a cross-over region where half-integer comet-like and trefoil-like defects coexist with their full-integer counterparts (Fig.~\ref{fig:pol}b). This is a hallmark of a state with mixed symmetry and indicates that the minimal model unifies both polar and nematic active turbulence. 

\begin{figure}[h]
\centering
  \includegraphics[width=0.75\linewidth]{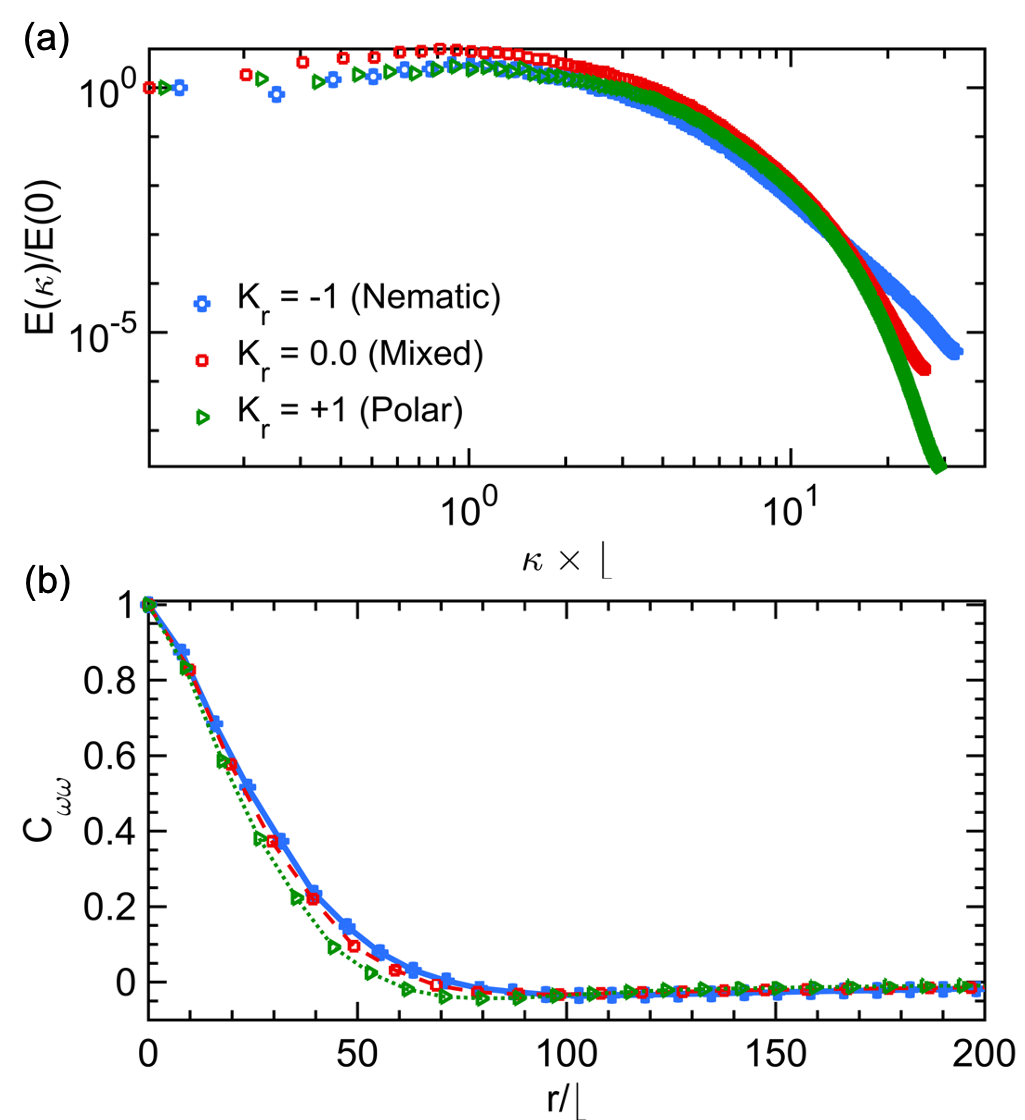}
  \caption{{\bf Turbulent flow characteristics.} (a) Kinetic energy spectrum and (b) vorticity-vorticity correlation function for the activity value $\bar{\zeta}=3.5$. The wave number $\kappa$ and length $r$ are normalised by the integral length $\lfloor=\int d\kappa~\left[E(\kappa)/\kappa\right]~/\int d\kappa~E(\kappa)$.}
  \label{fig:eng-corr}
\end{figure}

To quantify the emergence of mixed symmetry and transition from states with half-integer to full-integer topological defects, we measure the absolute value of the topological charge of the system $\langle|q|\rangle_{x,t}$, averaged over space and time, for varying dimensionless relative elasticity $K_r=(K_p-K)/(K_p+K)$. As such, if all defects are half-integer $\langle|q|\rangle_{x,t}=1/2$ and if all defects are polar $\langle|q|\rangle_{x,t}=1$. As evident from Fig.~\ref{fig:charge} upon increasing the relative elasticity $K_r$ the active system continuously crossovers from active nematic turbulence, characterized by apolar half-integer topological defects, to an active turbulence state with mixed polar and apolar symmetry, where half- and full-integer defects coexist, and to active polar turbulence, characterized by only polar full-integer topological defects. While, recent studies have suggested existence of universal scaling behaviors in active turbulence~\cite{alert2020universal}, the precise role of topological defects in setting turbulent flow characteristics in active fluids remains largely unexplored. We further characterize the flow properties of active turbulence states by measuring their associated kinetic energy spectrum and vorticity-vorticity correlation functions (Fig.~\ref{fig:eng-corr}). Interestingly, moving from active nematic turbulence to active polar turbulence is accompanied by an enhanced decay of the kinetic energy towards smaller scales and leads to lower vorticity correlation length, indicating that the nature of topological defects can have a marked impact on active turbulent flows.

The minimal model presented herein is the first to show how apolar, half-integer, topological defects can emerge in a continuum representation of polar active matter. Not only this presents a significant reduction in the complexity compared to the current active nematic formulation which is based on tensorial representation of the orientation, but also overcomes the limitation of active nematics by capturing the impact of the polarity of the particles. This is important because the majority of biological living systems such as bacterial suspensions or cellular monolayers for which half-integer topological defects have been identified~\cite{meacock2021bacteria,copenhagen2020topological,Saw2017,duclos2017topological,kawaguchi2017topological}, are composed of polar entities that continuously self-propel in the direction of polarity. Without the polar self-propelled forces, the current active nematic framework is basically modeling shakers that do not move but actively generate flows around themselves. 
Additionally, our findings provide a first characterization of the active turbulence with coexisting half- and full-integer defects. As such our approach unifies active polar and active nematics systems in one framework and provides testable predictions for observing states with mixed symmetry in the experiments.
Indeed, recent experiments on motility-assays have shown how modifying the interaction between the filaments can result in the coexistence of phases with polar and nematic symmetry in the absence of topological defects~\cite{huber2018emergence}. Future experiment could investigate the active turbulence generation in a dense system of polar filaments-motor protein mixtures and probe the co-existence of half- and full-integer topological defects within the active turbulence. Moreover, although recent experiments on epithelial monolayers have identified half-integer defects in the orientation field corresponding to the deformable shape of the cells~\cite{Saw2017,blanch2018turbulent,balasubramaniam2020nature}, earlier studies on epithelial monolayers have consistently reported the emergence of `rosette' structures, which closely resemble full-integer topological defects~\cite{blankenship2006multicellular,razzell2014recapitulation}, though their associated orientation field is yet to be fully characterized. We conjecture that epithelial monolayers could realize the active turbulence with mixed half- and full-integer topological defects and hope that our study triggers further experiments analyzing biological states with such mixed symmetry.
\section*{Acknowledgement and Funding}
AD acknowledges support from the Novo Nordisk Foundation (grant no. NNF18SA0035142), Villum Fonden (grant no. 29476), and funding from the European Union’s Horizon 2020 research and innovation program under the Marie Sklodowska-Curie grant agreement no. 847523 (INTERACTIONS).
\bibliographystyle{apsrev4-1}
\bibliography{references}
\end{document}